\documentclass[reprint,aps,pra,amsmath,amssymb,showpacs,nofootinbib]{revtex4-1}

\usepackage[pdftex]{graphicx}	
\usepackage{xcolor}
\graphicspath{ {figs/} }

\usepackage{hyperref}
\hypersetup{
	breaklinks=true,
	colorlinks=true,
	linkcolor=blue,
	citecolor=purple,
	pdfstartview={FitH},
	pdfview={FitH 0},
	pdfauthor={Jeffrey B. Parker},
	pdftitle={Nontrivial topology in the continuous spectrum of a magnetized plasma},
 }

\usepackage{bm}
\usepackage{eucal}		
\usepackage{mathrsfs}
\usepackage{mycommands}   

\newcommand{\Alfven}{Alfv\'en}

\newcommand{\unitgreek}[1]{\bm{\hat{#1}}}

\providecommand{\eqnref}{}
\renewcommand{\eqnref}[1]{Eq.~\eqref{#1}}

\providecommand{\defineas}{}
\renewcommand{\defineas}{\doteq}		

\newcommand{\vk}{\v{k}}
\usepackage{braket}
\newcommand{\sgn}{\operatorname{sgn}}

\begin{document}

\title{Nontrivial topology in the continuous spectrum of a magnetized plasma}

\author{Jeffrey B.~Parker}
\email{jeff.parker@wisc.edu}
\affiliation{Department of Physics, University of Wisconsin-Madison, Madison, Wisconsin 53706, USA}
\affiliation{Lawrence Livermore National Laboratory, Livermore, CA 94550, USA}

\author{J. W. Burby}
\affiliation{Los Alamos National Laboratory, Los Alamos, NM 87545, USA}

\author{J. B. Marston}
\affiliation{Brown Theoretical Physics Center and Department of Physics, Brown University, Providence, RI 02912-1843, USA}

\author{Steven M. Tobias}
\affiliation{Department of Applied Mathematics, University of Leeds, Leeds, LS2 9JT, United Kingdom}

\begin{abstract}
Classification of matter through topological phases and topological edge states between distinct materials has been a subject of great interest recently.  While lattices have been the main setting for these studies, a relatively unexplored realm for this physics is that of continuum fluids.  In the typical case of a fluid model with a point spectrum, nontrivial topology and associated edge modes have been observed previously.  However, another possibility is that a continuous spectrum can coexist with the point spectrum.  Here we demonstrate that a fluid plasma model can harbor nontrivial topology within its continuous spectrum, and that there are boundary modes at the interface between topologically distinct regions.  We consider the ideal magnetohydrodynamics (MHD) model.  In the presence of magnetic shear, we find nontrivial topology in the {\Alfven} continuum.  For strong shear, the Chern number is $\pm 1$, depending on the sign of the shear.  If the magnetic shear changes sign within the plasma, a topological phase transition occurs, and bulk-boundary correspondence then suggests a mode localized to the layer of zero magnetic shear.  We confirm the existence of this mode numerically.  Moreover, this reversed-shear {\Alfven} eigenmode (RSAE) is well known within magnetic fusion as it has been observed in several tokamaks.  In examining how the MHD model might be regularized at small scales, we also consider the electron limit of Hall MHD.  We show that the whistler band, which plays an important role in planetary ionospheres, has nontrivial topology.  More broadly, this work raises the possibility that fusion devices could be carefully tailored to produce other topological states with potentially useful behavior.
\end{abstract}

\maketitle

\section{Introduction}
Recent discoveries have demonstrated the fundamental importance of topology and topological phase to qualitative understanding of physical systems \cite{hasan:2010,bernevig:book,ozawa:2019}.  An important feature in topological understanding is the bulk-boundary correspondence principle: the boundary between two materials with differing topological phase has associated interface modes.  This effect has been observed in electronic \cite{hasan:2010}, photonic \cite{lu:2014}, acoustic \cite{yang:2015,he:2016}, and mechanical systems \cite{susstrunk:2015}.  The interface modes have attracted significant interest in enabling more efficient devices due to topological protection, the tendency to be robust against scattering in the presence of defects.  

Although most of the studies in this field have focused on systems with an underlying periodic lattice, the principles of topological phases can also be brought to bear on fluid systems \cite{delplace:2017,shankar:2017,perrot:2019,parker:2019gpp}.  Fluids represent a new frontier in topological phases of matter.  For instance, fluids are typically modeled with a continuum that ignores the discrete microscopic structure at small scales.  As a result, a homogeneous fluid possesses continuous translational symmetry, rather than the discrete translational symmetry of a lattice.  The unit cell size is effectively zero, and there is no finite Brillouin zone.

The mathematical equations describing fluids are generally nonlinear.  The standard procedure to analyze the wave spectrum is to linearize the equations about an equilibrium.  The resulting linear operator for the perturbation variables is the effective Hamiltonian.  While most real fluids vary in space, analysis of bulk topology is conveniently done by assuming a homogeneous system, which facilitates a Fourier transform and the calculation of the spectrum.

Here, we demonstrate that a topological phase is found in a novel setting, in the \emph{continuous} spectrum of an \emph{inhomogeneous} fluid.  A continuous spectrum is often associated with improper eigenfunctions that are not square-integrable.  We study ideal magnetohydrodynamics (MHD), a model describing the large-scale behavior of a magnetized plasma, or alternatively, a perfectly conducting fluid.

\begin{figure*}[!t]
	\centering
	\includegraphics{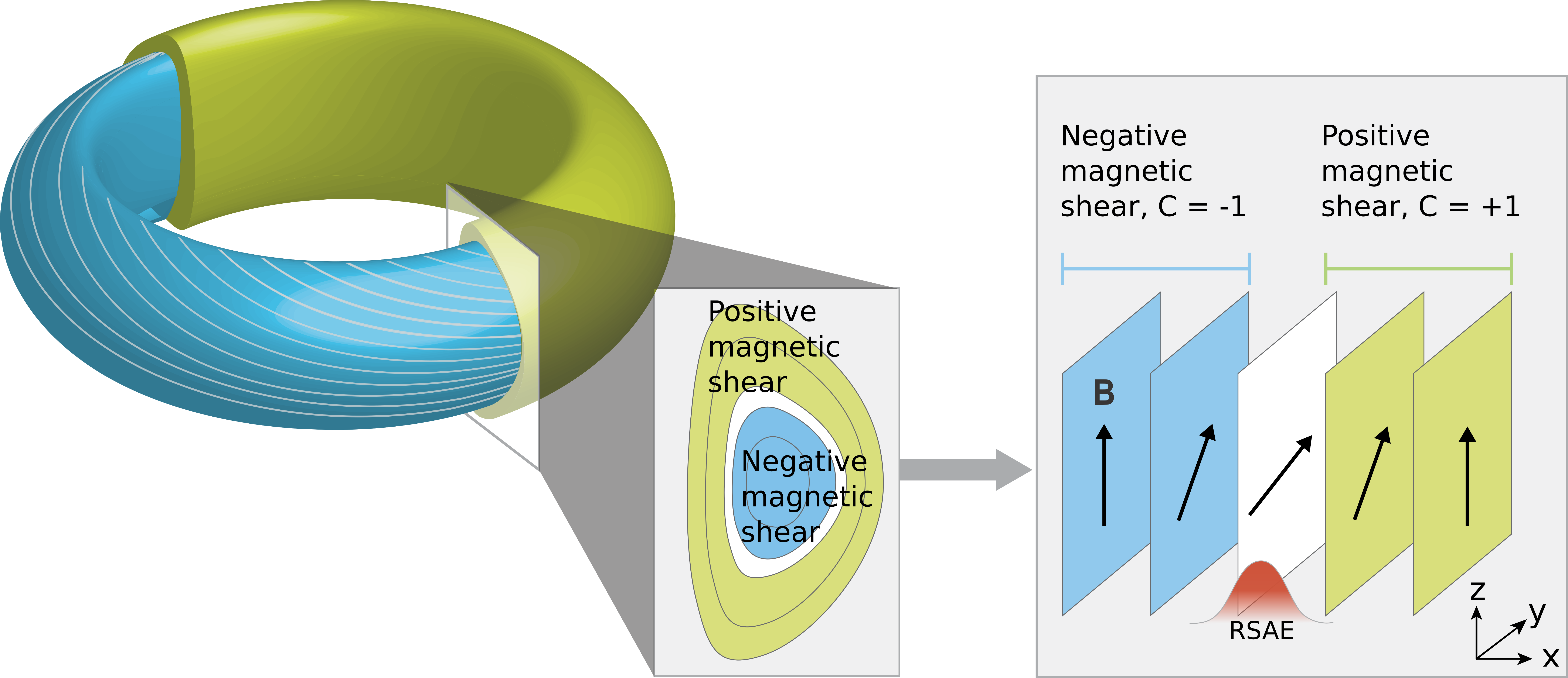}
	\caption{Left: Toroidal geometry of a tokamak along with a poloidal cross section showing several nested magnetic flux surfaces.  Right: Corresponding idealized slab model.  The blue, white, and light green colors indicate regions of negative, zero, and positive magnetic shear, respectively.  Each magnetic surface is characterized by a Chern number depending on its magnetic shear.  Surfaces of negative and positive shear are characterized by opposite Chern numbers $C$ and hence differing topologies.  The RSAE is localized to the zero-shear layer where a topological phase transition occurs.}
	\label{fig:schematic}
\end{figure*}

An inhomogeneous plasma gives rise to a continuous spectrum in ideal MHD \cite{pridmorebrown:1966,uberoi:1972,tataronis:1973,grossman:1973,tataronis:1975}.  It is within the {\Alfven} continuum spectrum that we find nontrivial topology.  An interesting complication is that eigenfunctions associated with the continuous spectrum are singular and non-square-integrable.  To deal with the singular eigenfunctions, we use a regularization parameter, compute the Chern number, and then take the limit of the regularization parameter tending to zero.  With this procedure, we can extract a nonzero Chern number.  The Chern number depends on the local magnetic shear, which is the rate at which the pitch angle of the magnetic field vector changes in space.

A cartoon schematic is depicted on the right side of Fig.~\ref{fig:schematic}.  Imagine a set of stacked, infinite planes, one for each position $x$.  Each plane is a magnetic surface, a two-dimensional surface in which a magnetic field line wholly lies.  Each magnetic surface is associated with a Chern number, which depends on the sign of the magnetic shear.  While the magnetic shear can vary continuously across different surfaces, the Chern number of each surface remains fixed as long as the shear does not change sign.  If the magnetic shear does change sign within the plasma, then the surfaces across the reversal layer have opposite Chern numbers.  A topological phase transition occurs across the shear-reversal layer.  Bulk-boundary correspondence suggests a discrete mode will exist, localized to the interface.  Indeed, we find an edge mode, a reversed-shear {\Alfven} eigenmode (RSAE), and numerically confirm its existence.

In fact, the RSAE is already well known within the magnetic fusion community.  In a toroidal magnetic confinement device such as a tokamak, the stacked layer of planes from the slab model is replaced by a set of nested torii that form concentric magnetic surfaces, shown on the left side of Fig.~\ref{fig:schematic}.  Some stable plasma equilibria can exist in which the magnetic shear vanishes on one of the surfaces.  RSAEs \cite{berk:2001,breizman:2003, heidbrink:2008, zonca:2002} have been observed in multiple tokamaks \cite{kusama:1998,sharapov:2002,nazikian:2003,sharapov:2004,snipes:2005,takechi:2005,vanzeeland:2006,fredrickson:2007,zhang:2018} and at least one stellarator \cite{toi:2010}.  The RSAEs are radially localized to regions of zero magnetic shear.  They are an MHD-type mode, and can be excited by kinetic effects such as resonance with energetic particles.  A concern for future fusion devices such as ITER is how the interaction between particles and {\Alfven} eigenmodes may limit reactor performance \cite{sigmar:1992, duong:1993,garciamunoz:2010,pace:2011}.  On the other hand, RSAEs have also proven useful for MHD spectroscopy \cite{sharapov:2001}, as they can be reliably detected, with an oscillation frequency close to that predicted by linear theory \cite{kramer:2006,deng:2012,chen:2013}.  Detection of the RSAEs can provide information about the magnetic field and safety factor that may be difficult to obtain by other means.

While RSAEs in tokamaks are known and have been both experimentally observed and theoretically studied and simulated, what is new here is the connection to topological physics.  The slab model of ideal MHD shares many features with the mathematical description in a torus or cylinder.  Most importantly, there exists a continuous spectrum which is determined by the same physics of where the frequency is equal to the local {\Alfven} frequency.  Hence, we suggest that the topological classification of magnetic surfaces could apply also to the torus or cylinder, with dependence on the sign of the magnetic shear, and that like the slab-RSAE, the RSAE of tokamaks and stellarators arises as the edge mode due to a topological phase transition within the plasma.  There are novel potential implications for tokamaks, as these results suggest that RSAEs may be topologically robust in the presence of three dimensional perturbations, such as magnetic islands.

These results may have broader implications for understanding of solar physics, where {\Alfven} waves can play an important role \cite{mathioudakis:2013, belien:1996}.  For instance the presence or absence of topological modes can yield information about stellar parameters \cite{perrot:2019}.

The outline for the rest of this paper is as follows.  In Sec.~\ref{sec:linearidealmhd}, we introduce linear ideal MHD, the continuous spectrum of the {\Alfven} continuum, and the improper eigenfunctions.  In Sec.~\ref{sec:topologyalfvencontinuum}, we calculate the Chern number of a magnetic surface and numerically confirm the existence of RSAE modes at a location where the Chern number changes sign.  In Sec.~\ref{sec:TRsymmetry}, we consider an apparent paradox regarding time-reversal symmetry in ideal MHD.  In Sec.~\ref{sec:hallmhd}, we consider the more detailed model of Hall MHD, which may provide a natural, physics-based regularization to MHD at small scales.  Finally, our conclusion is given in Sec.~\ref{sec:discussion}.

\section{Linear ideal MHD}
\label{sec:linearidealmhd}
For a collisional plasma, MHD can be rigorously derived as the large-spatial-scale, slow-time-scale description through a Chapman--Enskog procedure.  The core of a tokamak plasma is extremely hot and collisionless and a hydrodynamic description is not strictly valid, and yet MHD has proven to be a useful model for large-scale plasma behavior.  In particular, ideal MHD, which neglects transport coefficients like viscosity and resistivity, contains the basic ingredients for understanding {\Alfven} eigenmodes.  The basic equations of MHD are given in Appendix~\ref{app:idealmhd}.

\subsection{Inhomogeneous slab}
Here, we consider  a slab model in ideal MHD equilibrium, where equilibrium quantities depend only on $x$.  The equilibrium is assumed static, with no flow.  The plasma is homogeneous and infinite in $y$ and $z$.   The acceleration due to gravity is $-\hat{g}\unit{x}$, where $\hat{g}$ is a constant.  The gravitational force here plays a role fulfilled by magnetic curvature in toroidal geometry.  The equilibrium magnetic field is $\v{B}(x) = B_y(x) \unit{y} + B_z(x) \unit{z}$.

\begin{figure}
	\centering
	\includegraphics[width=\columnwidth]{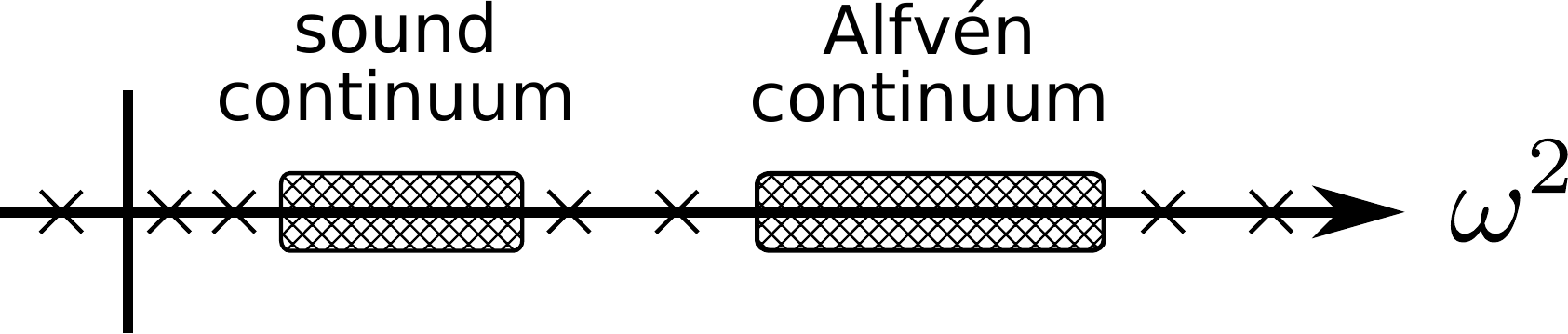}
	\caption{Schematic of the spectrum of ideal MHD of an inhomogeneous plasma.  For a given $k_y, k_z$, there are in general two sections of continuous spectrum, depicted in hatched regions; the sound continuum and the {\Alfven} continuum.  In the low-plasma-beta approximation, the sound continuum is filtered out of the equations.  The $\times$ symbols denote discrete eigenmodes, such as the RSAE.}
	\label{fig:schematic_mhd_spectrum}
\end{figure}

Perturbing about the equilibrium, we obtain the standard linearized ideal MHD formulation, $\r \partial^2 \bm{\xi} / \partial t^2 = \v{F}(\bm{\xi})$, where $\bm{\xi}$ is the fluid displacement vector, $\r$ is the equilibrium mass density, and $\v{F}$ is the self-adjoint force operator.  After a Fourier transform in time, this problem has self-adjoint form with eigenvalue $\omega^2$.   The full linearized equations are given in Appendix~\ref{app:linearizedmhdslab}.  It can be shown that two separate continuous spectra exist, the sound continuum and the {\Alfven} continuum, as depicted in Fig.~\ref{fig:schematic_mhd_spectrum}.  In what follows, we use the low-plasma-beta limit to isolate the {\Alfven} continuum by neglecting the slow magnetosonic wave and the sound continuum.  The {\Alfven} continuum is associated with shear {\Alfven} waves propagating on a given magnetic surface at a frequency equal to the local {\Alfven} frequency $\w_A^2 = k_\parallel^2 v_A^2$, where $k_\parallel$ is the wavenumber parallel to the magnetic field.



Assuming that modes have dependence $\bm{\xi}(\v{x}) = \bm{\xi}(x) e^{i(k_y y + k_z z - \w t)}$, we obtain the wave equation for the three components of $\bm{\xi}$, taken here as $\xi(x) \defineas \bm{\xi} \cdot \unit{x}$, $\eta(x) \defineas i\bm{\xi} \cdot \unit{e}_\perp$, and $\z(x) \defineas i\bm{\xi} \cdot \unit{b}$, and we have defined $\unit{b}(x) \defineas \v{B}/B$ and $\unit{e}_\perp(x) \defineas \unit{b} \times \unit{x}$.  Here, the symbol $\defineas$ denotes a definition.  The wave equation is \cite{goedbloed:text}
	\begin{subequations}
	\label{spectral_matrix_equation}
	\begin{align}
		-\r \w^2 \xi &= (\r v_A^2 \xi')' - \r f^2 v_A^2 \xi + (\r g v_A^2 \eta)' + \r g \hat{g} \eta + \r f \hat{g} \z, \\
		-\r \w^2 \eta &= -\r g v_A^2 \xi' + \r g \hat{g} \xi - \r (f^2 + g^2) v_A^2 \eta, \\
		-\r \w^2 \z &= \r f \hat{g} \xi,
	\end{align}
	\end{subequations}
where a prime denotes derivative with respect to $x$, $f(x) \defineas (k_y B_y + k_z B_z) / B$ and $g(x) \defineas (k_y B_z - k_z B_y) / B$ are the parallel and the signed perpendicular wavenumbers, and $v_A^2(x) \defineas B^2 / \r$ is the {\Alfven} speed squared.  The permeability $\m_0$ has been set to 1.  A derivation of Eq.~\eref{spectral_matrix_equation} can be found in Appendix~\ref{app:linearizedmhdslab}.  One can solve for $\eta$ and $\zeta$ in terms of $\xi$, as
	\begin{align}
		\eta &= \frac{g}{\omega^2 - (f^2 + g^2) v_A^2} \bigl(v_A^2 \xi' - \hat{g} \xi \bigr), \label{eta_intermsof_xi} \\
		\z &= -\frac{f \hat{g}}{ \omega^2} \xi . \label{zeta_intermsof_xi}
	\end{align}
One then obtains a single equation for $\xi$ \cite{goedbloed:text},
	\begin{equation}
		\d{}{x} \left( P \d{\xi}{x} \right) - Q \xi = 0,
		\label{xi_differential_equation}
	\end{equation}
where	
	\begin{align}
		P(x) & \defineas \frac{\r v_A^2 \w^2 \bigl( \w^2 - \w_A^2 \bigr)}{D}, \label{xi_eqn_P} \\
		Q(x) &\defineas -\Biggl\{ \r(\w^2 - \w_A^2) + \r' \hat{g} - \frac{\r (f^2 + g^2) \hat{g}^2 (\w^2 - \w_A^2)}{D}  \notag \\
			& \qquad - \left[\frac{\r \hat{g} \w^2 (\w^2 - \w_A^2)}{D} \right]'  \Biggr\}, \label{xi_eqn_Q}  \\
		D(x) &\defineas \w^2 \bigl[ \w^2 - (f^2 + g^2) v_A^2 \bigr], \\
		\omega_A^2(x)  &\defineas f^2 v_A^2.
	\end{align}
Here, $\omega_A^2$ is the local {\Alfven} frequency.  For any $k_y$, $k_z$, and position $x_0$, there is a solution with frequency $\omega^2 = \omega_A^2(x_0, k_y, k_z)$, which leads to the {\Alfven} continuum.  At this $\w^2$ and $x_0$, $P=0$ and $x_0$ is a regular singular point of \eqnref{xi_differential_equation}.  

We are now going to restrict ourselves to considering a single magnetic surface, so we fix $x_0$.  Let $s \defineas x - x_0$; we shall consider only small $s$.  In this and the next section, we use a notation where the subscript $0$ denotes evaluation at $s=0$.  For instance, $(\w_A^2)'_0 = \partial (\w_A^2) / \partial s|_{s=0}$.  When $(\w_A^2)'_0 \neq 0$, we construct a Frobenius series expansion for $\xi$ about $s=0$.  The details are given in Appendix~\ref{app:frobenius}; the important results are quoted here.  The Frobenius series results in a repeated indicial exponent of zero, implying there are two (degenerate) solutions,
	\begin{align}
		\xi_1 &= u(s), \label{xi1_nonsingular} \\
		\xi_2 &= u(s) \ln s + v(s), \label{xi2_singular}
	\end{align}
where $u(s)$ and $v(s)$ have a Taylor series.  The logarithmic singularity of $\xi_2$ is square integrable, but $\eta_2$, which depends on $\xi_2'$, is not.  

In the Taylor series $u(s) = 1 + u_1 s + \cdots$, we compute the coefficient $u_1$:
	\begin{equation}
		u_1 = u_{1a} + \frac{\hat{g}}{v_{A0}^2},
		\label{frobenius_u1}
	\end{equation}
where
	\begin{align}
		u_{1a} &\defineas \frac{g_0^2 Q_{0a}} {\r_0 (\omega_A^2)'_0}, \\
		Q_{0a} &\defineas -\r_0' \hat{g}.
	\end{align}
We have split off $u_{1a}$ because it plays an important role while the other term in Eq.~\eref{frobenius_u1} will cancel out.

The $\ket{\psi} \defineas [\xi_2, \eta_2, \zeta_2]$ are the improper eigenfunctions.  The dispersion relation of any particular magnetic surface $x_0$ is given by $\w^2 = \w_A^2(x_0, k_y, k_z)$.  The nonsingular eigenfunctions $[\xi_1, \eta_1, \zeta_1]$ are not important for this calculation.  We shall see shortly that $(w_A^2)'_0$ depends on the magnetic shear.  Thus, the eigenfunction $\ket{\psi}$ contains dependence on the magnetic shear through the coefficient of the second term in the Taylor series for $u(s)$.

\subsection{Regularization}
In differential geometric terms, for fixed $x_0$ the improper eigenfunctions form a line bundle over the base manifold of $\vk$ space.  Since these eigenfunctions are not square integrable, there is no clear way to obtain a Berry connection from them, which requires an inner product.  Our procedure is first to \emph{regularize} the eigenfunctions with a parameter $\e$, giving rise to a regularized line bundle.  We then place an inner product on the regularized bundle and determine the Berry connection in the usual way.  We shall find that the integral of the Berry curvature of the regularized bundle converges to a value as $\e \to 0$, which we identify with the Chern number of the {\Alfven} continuum.  

We regularize the eigenfunctions by replacing $s$ with $s + i\e$, which is equivalent to a Plemelj form and related to dissipation.  This regularization has also been used in {\Alfven}-wave studies of plasma heating \cite{hasegawa:1974,chen:1974}.  We take $\e > 0$.  In principle, dissipation can lead to new types of instabilities, but we assume a stable equilibrium as commonly realized in experiments.  Using Eqs.~\eref{eta_intermsof_xi} and \eref{zeta_intermsof_xi}, and that $\w^2$ is equal to $\w_A^2 = f^2 v_A^2$, we obtain the regularized eigenfunctions
	\begin{gather}
		\xi_2(s) = u(s) \ln(s + i \e), \\
		\eta_2(s) =  -\frac{1}{g} \left[ \frac{u(s)}{s + i \e} + u'(s) \ln(s + i \e) \right] + \frac{\hat{g}}{g v_A^2} u(s) \ln(s + i \e), \label{eigenfunction_eta_full} \\
		\z_2(s) = -\frac{\hat{g}}{f v_A^2} u(s) \ln(s + i\e),
	\end{gather}
where we have neglected $v(s)$ because its contributions to the rest of the calculation are subdominant in $\e$.

We define an inner product
	\begin{equation}
		\braket{\psi | \phi} \defineas \int ds\, h_\e(s) \bigl[ \xi_\psi^* \xi_\phi  +  \eta_\psi^* \eta_\phi  +  \zeta_\psi^* \zeta_\phi \bigr],
		\label{innerproduct_definition}
	\end{equation}
with weight function $h_\e = \e^2 / [\pi ( s^2 + \e^4)]$.  This weight function has a smaller width than the regularized eigenfunctions, and so effectively acts as a delta function to pick out the behavior on the singular surface at $s=0$.

\subsection{Local Magnetic Model}
At this point, we take an explicit model for the magnetic field,
	\begin{equation}
		\v{B}(s) = B_0 [(1 + s/L_B) \unit{z} + s\unit{y} / L_s],
	\end{equation}
where $B_0$, $L_B$, and $L_s$ are constants.  The magnetic shear of the surface at $s=0$ is characterized by the length scale $L_s$.  Finite $L_s$ provides shear.  This local linear model is rather generic for a sheared field within the vicinity of any particular surface, as we may rotate our coordinate system so that $\unit{z}$ lies along $\v{B}(0)$.  Since the details of the magnetic field enter the calculation of the Chern number only through $(\w_A^2)'_0$, this simple-looking expression is actually sufficient to characterize any equilibrium magnetic field near a given magnetic surface with nonzero shear.

\begin{figure}
		\includegraphics[width=\columnwidth]{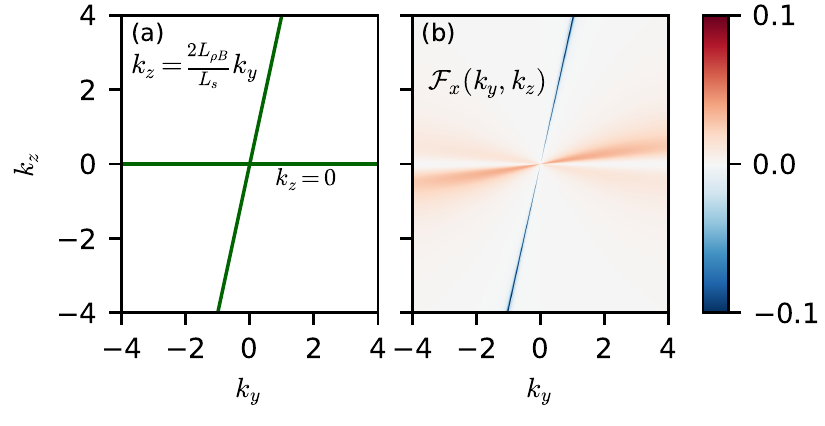}
		\caption{(a) Lines corresponding to zeros of $\bigl(\omega_A^2\bigr)'_0$ in the $(k_y,k_z)$ plane.  (b) Berry curvature $\mathcal{F}_x$.  The Berry curvature is localized near the lines where $\bigl(\omega_A^2\bigr)'_0$ is zero.  For this figure, $\epsilon = 0.05$, $L_{\r B} = 2$, $L_s=1$, $\hat{g}^2/v_{A0}^4 = 0.6$, and $Q_{0a}/B_0^2 = 0.5$.}
		\label{fig:berry_curvature}
	\end{figure}

For this model, $(\w_A^2)'_0$ is given by
	\begin{equation}
		\bigl(\w_A^2 \bigr)'_0 = \frac{B_0^2}{\r_0} \left( -\frac{1}{L_{\r B}} k_z^2  +  \frac{2}{L_s} k_y k_z \right),
	\end{equation}
where $1/L_{\r B} \defineas 1/L_\r - 2/L_B$ and $L_\r \defineas \r'_0 / \r_0$.  As mentioned earlier, the magnetic shear appears in the derivative $(\w_A^2 )'_0$, even though the shear does not appear in the local {\Alfven} frequency $(\w_A^2 )_0$ itself.  The quantity $(\w_A^2)'_0$ has non-isolated zeros in the $(k_y, k_z)$ plane along the lines $k_z=0$ and $k_z = 2L_{\r B} k_y/L_s$.  These zeros are important for the Chern integral and are shown in Figure \ref{fig:berry_curvature}(a).

\section{Topology of the {\Alfven} Continuum}
\label{sec:topologyalfvencontinuum}

\subsection{Berry curvature and Chern number}
\label{subsec:chernnumber}
In this section we calculate the Berry curvature and the Chern number.  We first must dispense with a technical complication: because of the 2-fold degeneracy of the eigenvalue, one must in general consider the non-Abelian connection $\mathcal{A}$ and curvature $\mathscr{F} = d\mathcal{A} + \mathcal{A} \wedge \mathcal{A}$.  However, the non-singular eigenfunction $[\xi_1, \eta_1, \zeta_1]$ has zero contribution to the first Chern form, which is proportional to $\operatorname{Tr}(\mathscr{F}) = \operatorname{Tr}(d\mathcal{A})$.  To see this, note that the non-singular eigenfunction may be written as pure real, by combining Eq.~\eref{xi1_nonsingular} with Eqs.~\eref{eta_intermsof_xi} and \eref{zeta_intermsof_xi}.  As a result, it does not change phase with $\vk$, and hence its diagonal entry in $\mathcal{A}$ vanishes.

Thus, we need only consider the regularized eigenfunction and its standard Berry connection, which for a normalized eigenfunction $\ket{\hat{\psi}}$ is given by
	\begin{equation}
		\v{A} = -\Im \bigl\langle \hat{\psi} | \nabla_{\v{k}} \hat{\psi} \bigr\rangle.
		\label{berryconnection_normalized}
	\end{equation}
It can be shown that $\bigl\langle \hat{\psi} | \nabla_{\v{k}} \hat{\psi} \bigr\rangle$ is purely imaginary; inserting the imaginary part operator in Eq.~\eref{berryconnection_normalized} can simplify the bookkeeping by allowing us to drop real terms along the way as all such terms must cancel out in the end.  It is convenient to work with a a non-normalized eigenfunction $\ket{\psi}$, for which an equivalent formula is
	\begin{equation}
		\v{A} = - \frac{ \Im \braket{\psi | \nabla_{\v{k}} \psi}}{ \braket{\psi | \psi}}.
	\end{equation}
Here, our parameter space consists of $(k_y, k_z)$, and $\nabla_{\vk} = \unit{y} \partial/\partial k_y + \unit{z} \partial / \partial k_z$.

The localization of the inner product weight function around $s = 0$ in Eq.~\eref{innerproduct_definition} implies we only need the eigenfunction evaluated at $s=0$, and we only require $u(0)=1$ and $u'(0) = u_1$ from the Frobenius series.  Moreover, the inner product also picks out $f(0) = k_z$ and $g(0) = k_y$.  For convenience, we state the regularized eigenfunctions evaluated at $s=0$:
	\begin{align}
		\xi_2(0) &= \ln \e, \label{regularized_eigenfunction_0_xi} \\
		\eta_2(0) &= -\frac{1}{k_y} \left[ u_{1a} \ln \e  -  i \left( \frac{1}{\e} - \frac{\pi u_{1a}}{2}\right) \right], \\
		\z_2(0) &= -\frac{\hat{g}}{k_z v_{A0}^2} \ln \e \label{regularized_eigenfunction_0_zeta},
	\end{align}
where we have used $\ln(i\e) = \ln \e + i\pi/2 \approx \ln \e$.  We have kept the imaginary piece  of the logarithm in $\eta_2$ for now because $u_{1a}$ depends on $\v{k}$ and becomes large at certain angles, though we shall see later that it does not affect the final result and could be neglected.  We also note that cancellation in $\eta_2$ has occurred between the third term and part of the second term of Eq.~\eref{eigenfunction_eta_full}, after one substitutes the first term in the series for $u$ and $u'$.  In Eqs.~\eref{regularized_eigenfunction_0_xi}--\eref{regularized_eigenfunction_0_zeta}, all of the dependence on the magnetic shear is buried in $u_{1a}$.  

A straightforward calculation yields
	\begin{align}
		\braket{\psi|\psi} &= \xi_2^*(0) \xi_2(0)  + \eta_2^*(0) \eta_2(0) + \z_2^*(0) \z_2(0) \\
			& =\frac{1}{k_y^2} w,
	\end{align}
where
	\begin{equation}
		w \defineas u_{1a}^2 \ln^2 \e + \left( \frac{1}{\e} - \frac{\pi u_{1a}}{2} \right)^2  +  \frac{k_y^2}{k_z^2} \frac{\hat{g}^2}{v_{A0}^4} \ln^2 \e  +  k_y^2 \ln^2 \e. \label{definition_of_w}
	\end{equation}
We then compute
	\begin{align}
		\Im \braket{\psi | \nabla_{\v{k}} \psi} &= \Im \bigl[ \xi_2^*(0) \nabla_{\v{k}} \xi_2(0) + \eta_2^*(0) \nabla_{\v{k}} \eta_2(0) \notag \\
			  & \qquad  + \zeta_2^*(0) \nabla_{\v{k}} \zeta_2(0)  \bigr].
	\end{align}
Since $\nabla_{\v{k}} \xi_2(0) = 0$, and $\z_2^*(0) \nabla_{\v{k}} \z_2(0)$ is real, the only contribution is from $\eta_2$ and is given by
	\begin{equation}
		\Im \bigl[ \eta_2^*(0) \nabla_{\v{k}} \eta_2(0) \bigr] = \frac{1}{k_y^2} \frac{\ln \e}{\e} \nabla_{\v{k}} u_{1a}.
	\end{equation}

Hence, the Berry connection is
	\begin{equation} 
		\v{A} = -\frac{\ln \e}{\e} \frac{1}{w} \nabla_\vk u_{1a}.
	\end{equation}	
	
The Berry connection determines the Berry curvature, given by
	\begin{equation}
		\bm{\mathcal{F}} = \nabla_\vk \times \v{A} = \frac{\ln \e}{\e} \frac{1}{w^2} \nabla_\vk w \times \nabla_\vk u_{1a}.
	\end{equation} 
It is convenient to evaluate this expression in polar coordinates.  Let $k_y = k\cos\vp$ and $k_z = k\sin\vp$. Noting that $u_{1a}$ depends only on $\vp$ and not on $k$, we find
	\begin{equation}
		\nabla_{\v{k}} u_{1a} = -\frac{\unitgreek{\vp}}{k} \frac{Q_{0a}}{B_0^2} \frac{\frac{2}{L_s} \cos^2\vp - \frac{2}{L_{\r B}} \cos\vp \sin\vp } {\bigl( \frac{2}{L_s} \cos\vp \sin\vp - \frac{1}{L_{\r B}} \sin^2\vp \bigr)^2}.
	\end{equation}

Most of the terms in $\nabla_{\v{k}} w$ do not contribute to $\bm{\mathcal{F}}$.  The first two terms on the right-hand side of Eq.~\eref{definition_of_w} have no contribution because they yield terms proportional to $\nabla_{\v{k}} u_{1a}$.  The third term also does not contribute because it only depends on $\vp$ and hence, like $\nabla_{\v{k}} u_{1a}$ its gradient is in the $\unitgreek{\vp}$ direction.  Only the last term on the right-hand side of Eq.~\eref{definition_of_w} contributes.   We obtain
	\begin{equation}
		\bm{\mathcal{F}} = -\unit{x} \frac{\ln^3 \e}{\e} \frac{2 Q_{0a}}{B_0^2} \frac{\cos^2 \vp}{w^2} \frac{ \frac{2}{L_s} \cos^2 \vp - \frac{2}{L_{\r B}} \cos\vp \sin\vp}{\Bigl( \frac{2}{L_s} \cos\vp \sin\vp - \frac{1}{L_{\r B}} \sin^2\vp \Bigr)^2}.
		\label{Berrycurvature}
	\end{equation}
Since $\mathcal{F}_x$ is proportional to $Q_{0a}$, the Berry curvature vanishes in the absence of gravity.  The Berry curvature is shown in Fig.~\ref{fig:berry_curvature}(b). It is localized near the lines where $(\w_A^2)'_0 = 0$, but the Berry curvature itself does not become singular for finite $\e$.

We compute the Chern number by integrating over the $(k_y, k_z)$ plane as 
	\begin{equation}
	C = \frac{1}{2\pi} \int dk_y\, dk_z\, \mathcal{F}_x.
	\label{Chern_integral}
	\end{equation}
The integral is evaluated in the $\e \to 0$ limit in Appendix~\ref{app:integral}.  The result is
	\begin{equation}
		C = -\sgn(\r_0' \hat{g} L_s) \left(1 - \frac{1}{\sqrt{1 + \frac{4 L_\r^2}{L_s^2}}} \right).
		\label{Chern_result}
	\end{equation}
Remarkably, the $\e$ dependence has canceled out, leaving a finite result.  This ``Chern number'' is in general not an integer.  A noninteger integral of the Berry curvature has been observed in other continuum fluid models to be the result of an insufficiently smooth linear operator at small scales, leading to the inability to compactify the space \cite{silveirinha:2015}.  This suggests that regularization of the ideal MHD linear operator at small scales could restore a proper integer Chern number.  In the limit of strong magnetic shear, $L_s \ll L_\r$, the second term in the parentheses vanishes and $C = -\sgn(\r_0' \hat{g} L_s)$.  In the opposite case of weak magnetic shear, $L_s \gg L_\r$, \eqnref{Chern_result} reduces to the topologically trivial $C=0$. 

As additional support for focusing on the first term in parentheses in \eqnref{Chern_result}, we observe that a Chern number should not change under smooth variations of the inner product.  Suppose instead of \eqnref{innerproduct_definition}, we used a modified inner product
	\begin{equation}
		\braket{\psi | \phi}_{\text{mod.}} = \int ds\, h_\e(s) \bigl[ \a \xi_\psi^* \xi_\phi  +  \b \eta_\psi^* \eta_\phi  +  \zeta_\psi^* \zeta_\phi \bigr],
	\end{equation}
where $\a$ and $\b$ are constants.  The result for the Chern integral is then modified to be
	\begin{equation}
		C_{\text{mod.}} = -\sgn\bigl(\r_0' \hat{g} L_s \bigr) \left( 1 - \frac{1}{\sqrt{1 + \frac{4 L_\r^2}{\b L_s^2}}} \right).
	\end{equation}
Only the second term is affected compared with Eq.~\eref{Chern_result}.  Hence, the fact that the result should be independent of $\a$ and $\b$ suggests the relative importance of the first term.   

The Chern number depends on the sign of the magnetic shear.  Therefore, the bulk-boundary correspondence principle implies that a mode will exist localized to a zero-shear layer that separates topologically distinct regions of positive and negative magnetic shear.  That this mode indeed exists is shown in the next section.

Previous analyses of \eqnref{xi_differential_equation} in the vicinity of a maximum of $\w_A^2$ have revealed a condition for an accumulation point of the discrete spectrum to occur \cite{goedbloed:text}.  Let $x_*$ be the location where $\w_A^2(x, k_y, k_z)$ reaches a maximum for given $k_y$, $k_z$.  Such a maximum may be created by a minimum in the slab analog of the safety factor, $B_z(x) / B_y(x)$.  The condition for the maximum of $\w_A^2$ to be an accumulation point is $2(k_y^2 + k_z^2) Q_* / \r_* |(\w_A^2)''_*|  > 1/4$, where $Q_* = -\r'_* \hat{g}$ and a $*$ subscript denotes evaluation at $x_*$.  This condition depends explicitly on the wave vector $(k_y, k_z)$.  A similar condition arises in a tokamak, with $Q_*$ replaced by contributions due to energetic particles and toroidicity \cite{berk:2001,breizman:2003}.

In contrast, the topological characterization here provides a different existence condition for interface modes, which apparently only requires nonzero $\r' \hat{g}$ and makes no reference to wave vectors.  Furthermore, in contrast to the local analysis just described, this conclusion is reached only by analyzing the {\Alfven} continuum \emph{away} from the region of zero shear, and using the bulk-boundary correspondence principle to infer the existence of a mode if a zero-shear layer exists.  

Before moving on from this section, we consider an additional question.  Because the RSAE can also be described within reduced MHD, it is natural to ask if the same topology is found in reduced MHD.  In Appendix~\ref{app:reducedmhd}, we show that in contrast to MHD, within reduced MHD the {\Alfven} continuum has trivial topology.  This result raises an apparent paradox: if the RSAE is associated with the nontrivial topology of the Alfv\'{e}n continuum in MHD, how can it be reconciled that the RSAE is found in reduced MHD?

We resolve the apparent paradox as follows.  The topology of the vector bundle is a global property of the eigenfunction on the entire $\vk$ space.  As we discuss in Sec.~\ref{sec:compactness}, the computed topology can be sensitive to details that do not affect the physical modes of interest in some local region of $\vk$ space.  Hence, although reduced MHD still contains the local-in-$\vk$-space analytic ingredients for the RSAE, approximating MHD down to reduced MHD alters the global topological structure of the {\Alfven} continuum.  As shown in Appendix~\ref{app:reducedmhd}, the approximation causing the difference in global topology is the neglect of perturbations parallel to the equilibrium magnetic field---a severe alteration to the description of the eigenfunction.

\subsection{Numerical calculation of RSAE interface mode}
We confirm the existence of the RSAE at the region of zero shear by numerically solving Eq.~\eref{spectral_matrix_equation} directly as an eigenvalue equation using the code Dedalus \cite{burns:2019}.  We use $B_z = 5$, $B_y(x) = 1.52 x - 1.216 x^2$, $\r(x) = \exp(-x/20)$, and $\hat{g}=75$.  The boundary conditions are $\xi(0) = \xi(1) = 0$.

\begin{figure}
		\includegraphics[width=\columnwidth]{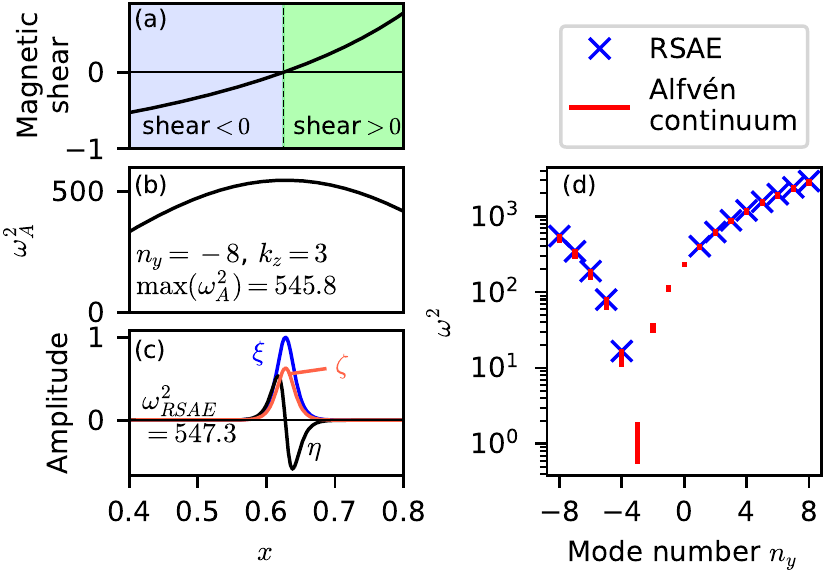}
		\caption{Numerical example confirming the existence of the RSAE where the magnetic shear changes sign.  (a) Magnetic shear $(x/q) dq/dx$, where $q = B_z(x) / B_y(x)$ is the safety factor.  (b) {\Alfven} continuum $\w_A^2(x)$ and (c) RSAE eigenfunction at $n_y = -8$, with $k_y = 10n_y$ and $k_z=3$.  (d) RSAE spectrum as a function of $n_y$.  The fundamental mode is shown, which has eigenvalue $\w^2$ larger than the maximum of $\w_A^2$.  Also shown is the {\Alfven} continuum $\w_A^2$ in the interval $0.5 < x <0.75$ (outside this interval, the RSAE has little amplitude and would not interact much with the continuum).}
		\label{fig:slabRSAEspectrum}
	\end{figure}

Figure \ref{fig:slabRSAEspectrum} shows the numerical solution; see the caption for details.  The numerical solution confirms that the RSAE is an eigenmode and exists at the layer where the magnetic shear vanishes.  For this plot, we discretize $k_y = n_y \D k_y$ with $\D k_y = 10$ and integer $n_y$, and we hold constant $k_z = 3$.  We observe that given $(k_y, k_z)$, the frequency of the RSAE lies above the maximum of the {\Alfven} continuum frequency, and thus the factor $P(x)$ in Eq.~\ref{xi_eqn_P} is never zero, so the singularity is avoided.  The frequency of the fundamental mode is shown.  Solutions with additional spatial nodes may sometimes be found, and have frequency closer to the maximum of the continuum.  For $-3 \le n_y \le 0$, no RSAE is found.  In that range, $\omega_A^2$ inverts and forms a local minimum at $x \approx 0.625$ instead of a local maximum.
 
\subsection{Compactness}
\label{sec:compactness}
Because the Chern number in Eq.~\eref{Chern_result} need not be an integer, a brief discussion of compactness is merited.  The Chern theorem states that for a compact manifold, the integral of the Berry curvature is an integer.  The wavevector space of systems based on periodic lattices is also periodic and is therefore compact.  In contrast, the continuum structure of fluid models means that wavevector space extends to infinity and is not compact.  That is, because the unit cell size is zero in a fluid, there is no finite Brillouin zone.  Because the wavevector space is not compact, the Chern theorem does not apply and one is not guaranteed an integer result for the integral of the Berry curvature.

Various ways for dealing with the lack of compactness in continuum models have been discussed \cite{silveirinha:2015,tauber:2019,souslov:2019}.  If the integral of the Berry curvature turns out to not be an integer, one can perform a sort of regularization of the problem at small scales, which allows the space to be compactified, for instance, to the Riemann sphere.  One physically motivated regularization is based on the idea that at sufficiently small scales such as lengths shorter than the interparticle spacing, the continuum description is no longer physically valid \cite{silveirinha:2015}.  An \emph{ad hoc} regularization to the fluid system at arbitrarily small scales can emulate this behavior and restore compactness and an integer Berry curvature integral.

On the other hand, compactness is a property that simplifies the discussion, but may not matter for practical purposes.  In particular, an \emph{ad hoc} fix at small scales that enables compactification may not necessarily affect whether an edge mode exists at an interface between topologically distinct regions.  For example, in Ref.~\cite{parker:2019gpp}, interface modes at the boundary between gaseous plasma and vacuum can be found regardless of whether a regularization is applied at arbitrarily small scales.

From a theoretical viewpoint, an index theorem relating an analytical index and topological index continues to hold for non-compact manifolds, but there will be additional boundary terms that must be accounted for.  See, e.g., Ref.~\cite{eguchi:1980}.  From this point of view, noninteger integrals of the Berry curvature are likely related to boundary data arising from the infinite wave vectors.

The application of these ideas to understanding the topology of the continuous spectrum of MHD in more detail will be undertaken in future work.  Most likely, a complete and rigorous understanding of this point will require more microscopic physics to be brought in.  A partial investigation is explored in Sec.~\ref{sec:hallmhd}.

\section{Time-Reversal Symmetry in MHD}
\label{sec:TRsymmetry}
If one considers time-reversal symmetry in ideal MHD, a counterintuitive mathematical behavior arises.  Consider the standard time-reversal (TR) transformation $T$, which acts in the following way:
	\begin{equation}
		T: \quad t \mapsto -t, \quad \v{v} \mapsto -\v{v}, \quad \v{B} \mapsto -\v{B},
	\end{equation}
and all other variables, including $\v{x}$, $p$, and $\r$, remain unchanged.  Under $T$, the nonlinear ideal MHD equations \eref{app:MHDcontinuity}--\eref{app:MHDinduction} are invariant.

In many situations of interest, an external magnetic field is applied, whose generating current should be regarded as extrinsic to the system under study.  The magnetic field can be written as $\v{B} + \v{B}_a$, where $\v{B}_a$ is the applied field.  Under $T$, $\v{B}$ changes sign and $\v{B}_a$ does not change sign.  Hence, in the presence of an externally applied magnetic field, ideal MHD is not invariant under time-reversal transformation $T$.

We can, however, define a modified TR transformation $T'$ in which the magnetic field does not change sign, and under which ideal MHD is invariant.  In actuality $T'$ is just $CT$, the composition of charge conjugation and time reversal, since charge conjugation acts to reverse the sign of $\v{B}$.  $T'$ acts as
	\begin{equation}
		T': \quad	t \mapsto -t, \quad \v{v} \mapsto -\v{v},
	\end{equation}
and $\v{B}$, $\v{x}$, $p$, and $\r$ are unchanged.  Then Eqs.~ \eref{app:MHDcontinuity}--\eref{app:MHDinduction} are invariant under $T'$, even in the presence of an externally applied magnetic field.

Physically, MHD can be invariant under $T'$ because electrons are not explicitly modeled within MHD, and the MHD current is therefore decoupled from the MHD fluid velocity.  Unlike in classical electrodynamics, in which the current is proportional to the particle velocity, in MHD the current and the magnetic field need not transform like the fluid velocity.

We turn from the nonlinear ideal MHD equations to the linearized equations about a static equilibrium.  The linearized equations are given in Eqs.~ \eref{app:linearizedMHDcontinuity}--\eref{app:linearizedMHDinduction}.  The equilibrium magnetic field $\v{B}_0$ behaves like an external field with respect to the perturbation variables.  Hence, under a transformation $T$ in which $t$, $\v{v}_1$, and $\v{B}_1$ change sign, the linearized MHD equations are not invariant.  But under $T'$ in which $t$ and $\v{v}_1$ change sign but $\v{B}_1$ does not, linearized MHD is invariant.  Linearized ideal MHD can also be recast in terms of the fluid displacement variable $\bm{\xi}$, as given in Eq.~\eref{app:idealmhd_xiequation1}.  This equation is also invariant under $T'$, in which $t$ changes sign but neither $\v{x}$ nor $\bm{\xi}$ changes sign.

The existence of a TR transformation $T'$ under which the linearized ideal MHD equations are invariant suggests a paradox.  A standard result is that TR invariance implies that the Berry curvature satisfies
	\begin{equation}
		\bm{\mathcal{F}}(-\v{k}) = -\bm{\mathcal{F}}(\v{k}).
	\end{equation}
Hence, the Chern number, which is the integral of $\bm{\mathcal{F}}$, would integrate to zero because the contribution of each $\v{k}$ would be canceled out by its partner wave vector $-\v{k}$.  We seem to have a contradiction.  Linearized ideal MHD is TR invariant, yet we have calculated a nonzero Chern number.

The resolution of the apparent paradox is our regularization of the problem.  First we note that we did not regularize the equations themselves, but rather the eigenfunction, in which we set $s \to s + i\e$.  This is effectively equivalent to adding a damping parameter for $\e > 0$, so we would expect this regularization to break TR invariance.  The regularized eigenfunction becomes $\xi = u(s) \ln(s + i\e) + v(s)$.  In Fourier space, time reversal can be effected by setting $\v{k} \to -\v{k}$ and applying complex conjugation.  Complex conjugation after regularization clearly breaks TR invariance of the eigenfunction.

The phenomenon of Landau damping provides an illuminating analog.  A Fourier transform of the collisionless Vlasov equation with real frequency leads to singularities at velocities $v = \omega /k$.  It has long been recognized that the more mathematically rigorous way to handle the problem is to approach it as an initial-value problem and apply a Laplace transform instead of a Fourier transform, which directly leads to the recovery of Landau damping \cite{landau:1946}.  However, even within a Fourier transform, it can be shown that a continuous spectrum exists, the so-called Case--Van Kampen modes, each of which is undamped \cite{kampen:1955,case:1959}.  Any physical mode must consist of an infinite number of such modes, and the phase mixing of different frequencies leads to Landau damping in an alternate guise.  To complete matters, there is a third way to recover Landau damping: we can return to the naive Fourier transform with singularities at $\w / k$.  If one inserts a regularization damping parameter tending to zero, denoted by $0^+$, into the frequency such that $\w \to \w - i0^+$, the singularity is eliminated and one recovers the Landau damping rate \cite{bellan:text}.  Landau damping occurs within a TR invariant system.

The ideal MHD spectrum is similar.  Linearized ideal MHD, like the collisionless Vlasov equation, is TR invariant and contains a continuous spectrum.  Any physical initial condition would consist of an infinite number of the continuum eigenfunctions, and phase mixing leads to a ``decay'' in time \cite{tataronis:1973,grossman:1973,tataronis:1975}.  Just as a damping parameter for regularization can be used in the Vlasov problem to obtain the correct answer for Landau damping, a regularization parameter within a Fourier transform formalism has been used within ideal MHD for similar purposes.

These analogs provide some motivation for using this type of TR-breaking regularization.  Moreover, the model of ideal MHD plus $0^+$ damping might be considered more physical than ideal MHD itself, so should perhaps be preferred if the two give different answers.  In the next section, we consider an alternate approach and use a model with additional microscopic electron physics compared to MHD that leads to a more natural breaking of TR symmetry.

\section{Hall MHD and Electron MHD}
\label{sec:hallmhd}
As discussed in Sec.~\ref{sec:compactness}, the non-integer Chern numbers found in Sec.~\ref{subsec:chernnumber} may result from insufficiently smooth small-scale behavior of the MHD model.  MHD is the theory of a conducting fluid and focuses on the ions, which carry almost all of the mass and momentum of a fluid.  MHD is a useful description of large scales, but does not provide the correct small-scale physics.  Hall MHD is an extension of MHD that incorporates additional electron dynamics and provides one possibility of resolving the smaller scales \cite{huba:2003}.  The purpose of this section is to explore whether Hall MHD provides a proper regularization of MHD to resolve the non-integer Chern numbers.  Tentatively, it appears the answer is yes.

Hall MHD is described by the same equations as MHD, Eqs.~\eref{app:MHDcontinuity}--\eref{app:MHDinduction}, except that Eq.~\eref{app:MHDinduction} is replaced by
	\begin{equation}
		\pd{\v{B}}{t} = \nabla \times (\v{v} \times \v{B}) - \nabla \times \left( \frac{\v{J}}{n_e e} \times \v{B} \right).
		\label{Hall_induction_equation}
	\end{equation}
where $\v{J} = \nabla \times \v{B}$ is the current density and $n_e$ is the electron density.  This equation  has the additional Hall term proportional to $\v{J} \times \v{B}$ compared to Eq.~\eref{app:MHDinduction}.  This equation can be obtained by combining the electron momentum equation (with electron inertia neglected) and Faraday's law \cite{huba:2003}.

The {\Alfven} continuum is resolved into discrete modes in Hall MHD \cite{ohsaki:2004,ito:2004}.  Mathematically, the {\Alfven} continuum appears in MHD because the highest-order derivative vanishes on surfaces where the frequency is equal to the local {\Alfven} frequency, as in Eq.~\eref{xi_differential_equation}.  The inclusion of Hall physics coupled to sound waves adds a higher-order derivative.  An alternative way to resolve the singularity is through inclusion of kinetic effects, where the shear {\Alfven} wave becomes the kinetic {\Alfven} wave at scales comparable to the ion gyroradius \cite{hasegawa:1976,lysak:1996,hollweg:1999}.

In Sec.~\ref{sec:TRsymmetry}, we noted a peculiarity of ideal MHD, that it is invariant with respect to a modified time-reversal transformation that does not change the sign of $\v{B}$.  Inclusion of electron physics into Hall MHD breaks this property.  The Hall induction equation \eref{Hall_induction_equation} is only invariant to the standard TR transformation in which $\v{B}$ changes sign, and it is \emph{not} invariant under the modified transformation $T'$ in which $\v{B}$ does not change sign.  Furthermore, in the presence of an applied magnetic field, there is no longer any TR symmetry in Hall MHD, which contrasts with the behavior in MHD.

To simplify our analysis, rather than deal with Hall MHD in full, we consider a limit known as electron MHD.  Electron MHD treats ions as a fixed stationary background and focuses on the physics provided by the electrons \cite{gordeev:1994,huba:2003} as well as the effects of breaking TR symmetry.  Hence, the physics is only valid for timescales shorter than ion times, and length scales shorter than ion lengths.  In a collisionless plasma, electron MHD is not valid for very small length scales, as the physical model breaks down, for example due to the presence of electron cyclotron orbits.  Our purpose here is not to describe the full, overall behavior of a plasma, but to examine the characteristics of the dispersion relation on scales smaller than MHD, and how that might affect the topology.

The electron MHD equation is given by Eq.~\eref{Hall_induction_equation} with fixed ions, $\v{v}=0$,
	\begin{align}
		\pd{\v{B}}{t} &= -\nabla \times \left( \frac{\v{J}}{n_e e} \times \v{B} \right) \notag \\
				&= -\frac{1}{n_e e} \nabla \times (\v{J} \times \v{B} )  +  \frac{1}{n_e^2 e} \nabla n_e \times (\v{J} \times \v{B}).
	\end{align}
The first term on the right-hand side of this equation describes whistler waves, and the second term describes Hall drift waves \cite{huba:2003}, which depends on nonzero $\nabla n_e$.

We will consider linear dynamics about a static plasma equilibrium.  As in Secs.~\ref{sec:linearidealmhd} and \ref{sec:topologyalfvencontinuum}, we consider a background magnetic field,
	\begin{equation}
		\v{B}_0 = B_0 \left(\unit{z} + \frac{x}{L_s} \unit{y} \right).
	\end{equation}
The current is then proportional to the magnetic shear,
	\begin{equation}
		\v{J}_0 = \frac{B_0}{L_s} \unit{z}.
	\end{equation}
Consider a plasma in force balance equilibrium.  Now, we suppose we are looking at a region of space confined within $|x| \ll L_s$.  This assumption enables us to set $x/L_s$ to zero, yielding $\v{B}_0 = B_0 \unit{z}$ and $\v{J}_0 = B_0 / L_s \unit{z}$.  Thus, we have made the standard local approximation of taking both $\v{B}_0$ and its gradient to be spatially uniform.  This procedure removes inhomogeneity in our specification of the problem, allowing us to use standard Fourier analysis.  Additionally, we see that the equilibrium is force-free within this approximation, $\v{J}_0 \times \v{B}_0 = 0$.  Note that magnetic shear is included in this homogeneous formulation.  That this can occur within electron MHD stands in contrast with the situation in MHD, where it is not possible to include magnetic shear in a formulation within a homogeneous fluid.

By considering only $x / L_s \ll 1$, we are effectively restricting ourselves to wave vectors with $k_x L_s \gg 1$.  Hence, we are zooming in on a particular magnetic surface, as we did in Sec.~\ref{sec:linearidealmhd}.

With $\nabla n_e = 0$, the linearized electron MHD equation is
	\begin{equation}
		\pd{\v{B}_1}{t} = -\frac{1}{n_e e} \nabla \times \bigl( \v{J}_1 \times \v{B}_0  +  \v{J}_0 \times \v{B}_1 \bigr).
	\end{equation}
Using Fourier analysis with $\partial_t \to -i\w$ and $\nabla \to i\v{k}$, this equation written out in Cartesian components becomes
	\begin{equation}
		\w \begin{pmatrix}
			B_{1x} \\ B_{1y} \\ B_{1z}
		\end{pmatrix}
		 =
		 H \begin{pmatrix}
			B_{1x} \\ B_{1y} \\ B_{1z}
		\end{pmatrix},
	\end{equation}
where the Hermitian linear operator $H$ is
	\begin{equation}
		H = \frac{v_A c}{\w_{pi}} \begin{bmatrix}
			\dfrac{k_z}{L_s} & -i k_z^2 & i k_y k_z \\[2ex]
			ik_z^2   &   \dfrac{k_z}{L_s}   &   -i k_x k_z \\[2ex]
			-i k_y k_z  &  i k_x k_z   &   \dfrac{k_z}{L_s}
		\end{bmatrix},
	\end{equation}
$v_A$ is the {\Alfven} speed, $c$ is the speed of light, $\w_{pi} = (n_i q_i^2 / \e_0 m_i)^{1/2}$ is the ion plasma frequency, and $c/\w_{pi}$ is the ion skin depth.

\begin{figure}
	\centering
	\includegraphics{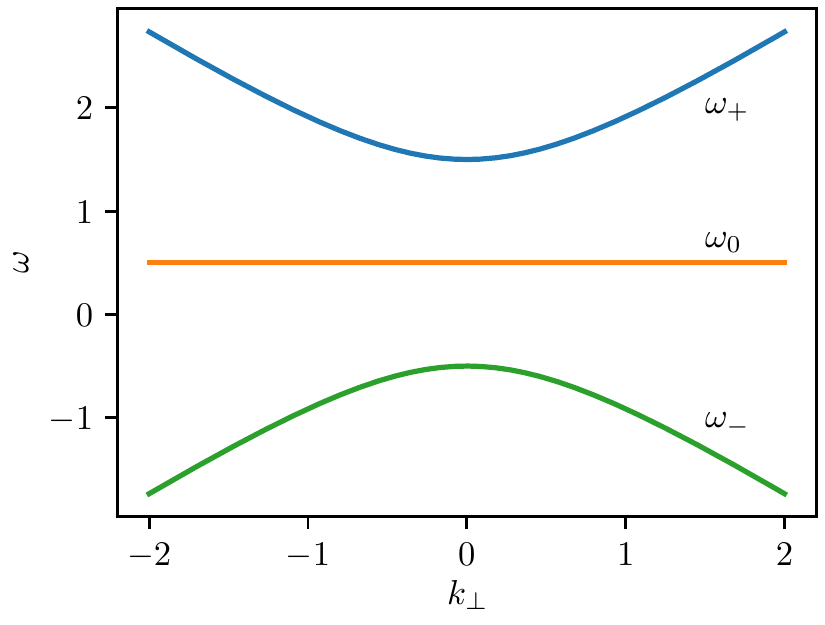}
	\caption{Dispersion relation of the three frequency bands in electron MHD as a function of $k_\perp = \pm\sqrt{k_x^2 + k_y^2}$.  For this figure, $v_A c/\w_{pi}=1$, $k_z = 1$, and $L_s=2$.}
	\label{fig:emhd_dispersion}
	\end{figure}

We proceed to find the eigenvalues and eigenvectors of $H$.  The magnetic shear appears only on the diagonal in the form $k_z/L_s$, and so gives a frequency shift.  The frequency roots are given by
	\begin{align}
		\w_0 &= \frac{v_A c}{\w_{pi}} \frac{k_z}{L_s}, \\
		\w_{\pm} &= \frac{v_A c}{\w_{pi}} \left( \frac{k_z}{L_s} \pm k_z k \right),
	\end{align}
where $k^2 = k_x^2 + k_y^2 + k_z^2$.  For $\v{k} \approx k_z \unit{z}$, we recover the whistler wave dispersion relation in $\w_\pm$, valid in the intermediate regime between the ion and electron cyclotron frequencies.  The frequency bands are shown in Fig.~\ref{fig:emhd_dispersion}.

The non-normalized eigenmode $\ket{\psi_0}$ corresponding to $\w_0$ is
	\begin{equation}
		\ket{\psi_0} = \begin{bmatrix} k_x \\ k_y \\ k_z \end{bmatrix}.
	\end{equation}
Similarly, the eigenmode $\ket{\psi_+}$ corresponding to $\w_+$ is
	\begin{equation}
		\ket{\psi_+} = \begin{bmatrix} k_x k_z - i k k_y \\  k_y k_z + i k k_x \\ -(k_x^2 + k_y^2) \end{bmatrix}.
		\label{hallmhd:psiplus1}
	\end{equation}
The eigenmode for $\w_-$ can be found similarly.
	
We analyze the topology of the frequency bands.  The analysis turns out to be similar to the physically distinct problem of shallow water waves \cite{delplace:2017}.  Because the components of $\ket{\psi_0}$ are all real and do not change phase with $\v{k}$, the middle branch of the dispersion curve is topologically trivial.  We next consider the upper frequency band; the lower band is analogous.  We consider a spherical surface in $\v{k}$ space, parameterized by the two angles $\th$ and $\p$ in the standard spherical coordinates with $k_x = k \sin\th \cos \p$, $k_y = k \sin \th \sin \p$, and $k_z = k \cos \th$.  The eigenvector $\ket{\psi_+}$ becomes, after normalizing,
	\begin{equation}
		\ket{\psi_+}(\th, \p) = \frac{1}{\sqrt{2}} \begin{bmatrix} 
			\cos\th \cos \p - i \sin \p \\
			\cos \th \sin \p + i \cos \p \\
			-\sin \th		
		 \end{bmatrix}.
	\end{equation}
This vector field contains a singularity at the north and south poles, which we momentarily ignore.  Using the gradient in spherical coordinates $\nabla_{\v{k}} = \unitgreek{\th} k^{-1} \partial_\th + \unitgreek{\p} (k\sin\th)^{-1} \partial_\p$, we have
	\begin{align}
		\nabla_{\v{k}} \ket{\psi_+} &= -\unitgreek{\th} \frac{1}{k} \begin{bmatrix} \sin\th \cos\p \\ \sin \th \sin \p \\ \cos \th \end{bmatrix} \notag \\
			&\quad   +  \unitgreek{\p} \frac{1}{k\sin\th} \begin{bmatrix} -\cos\th \sin\p  -  i\cos\p \\ \cos\th \cos\p  - i\sin\p \\ 0 \end{bmatrix}
	\end{align}
Hence, with the standard inner product, the Berry potential for this band is given by
	\begin{equation}
		\v{A}_+ = \frac{\cos\th}{k \sin \th} \unitgreek{\p}.
	\end{equation}
The Berry curvature is then given by
	\begin{equation}
		\bm{\mathcal{F}}_+ = -\frac{\unit{k}}{k^2}.
		\label{hallmhd:Berrycurvature1}
	\end{equation}
This is the field of a monopole located at the origin.  By the spherical symmetry of the Berry curvature, we see there are no problems at the poles.  Hence, any closed surface containing the origin has a Chern number of $-2$.

We can also consider the Chern number of the $(k_x, k_y)$ plane with $k_z$ fixed.  The details of the calculation are straightforward and are omitted.  On this surface, the Berry curvature is given by $\bm{\mathcal{F}_+} = -\unit{z} k_z / k^3$ [which is merely the $z$ component of the Berry curvature in Eq.~\eref{hallmhd:Berrycurvature1}], and the Chern number after integrating over the $(k_x,k_y)$ plane is $C = -\sgn(k_z)$.  This is consistent with the previous result, if we consider a closed cylindrical surface consisting of the $(k_x, k_y)$ planes at heights $\pm k_{z0}$ with radius tending to infinity.  The flux of Berry curvature through the sides of the cylinder is zero, and the total Chern number, with contributions from the planes at $\pm k_{z0}$, is $-1 -1 = -2$.  In a completely analogous way, we can calculate the Chern number of the $(k_y, k_z)$ plane with $k_x$ fixed to be $C = -\sgn(k_x)$.

We conclude that there are integer Chern numbers for all three bands of the electron MHD description.  The whistler band has nontrivial topology.  Interestingly, the Chern number is independent of the magnetic shear.  These findings provide evidence that inclusion of more detailed microscopic physics in the MHD calculation of Sec.~\ref{sec:topologyalfvencontinuum} could result in an integer Chern number.  We suggest a possibility for how this may work.  At the large scales where ion physics dominate, there is a nontrivial contribution to the Chern number that depends on the sign of the magnetic shear.  This is the piece derived above in Eq.~\eref{Chern_result}.  The non-integer piece of that Chern number may arise due to compactification issues at small scales in MHD.  But instead of using MHD at infinite $\v{k}$, we can consider a more realistic model.  The electron MHD model here, which is valid at intermediate $\v{k}$ but not strictly valid at very small $\v{k}$ (where ion physics reigns) or very large $\v{k}$ in collisionless regimes (where electron cyclotron orbits matter), shows that instead of a non-integer piece, one gets an integer contribution, and in particular one that does not depend on the sign of the magnetic shear.  

This description, while attractive, comes from considering the ion scales and the electron scales separately.  To be more rigorous, it would be valuable to use a single mathematical description that captures ion and electron scales simultaneously.  This is outside the scope of the present work but provides a direction for the future.

We finish this section by returning to our earlier assumption of neglecting $\nabla n_e$.  If we include it in our analysis, taking $\nabla n_e = \unit{x} n_e/L_n$, we obtain a new linear operator in place of $H$,
	\begin{equation}
		H' = \frac{v_A c}{\w_{pi}} \begin{bmatrix}
			\frac{k_z}{L_s} & -i k_z^2 & i k_y k_z \\[1ex]
			ik_z^2   &   \frac{k_z}{L_s}   &   -i k_x k_z \\[1ex]
			-i k_y k_z + i \frac{1}{L_n L_s}  &  i k_x k_z  - \frac{k_z}{L_n}   &   \frac{k_z}{L_s} + \frac{k_y}{L_n}
		\end{bmatrix}.
	\end{equation}
Hence, for nonzero $\nabla n_e$ the linear operator is no longer Hermitian.  One could then apply the theory of topology for non-Hermitian Hamiltonians \cite{leykam:2017,shen:2018}, which is another potential direction for future research.

\section{Discussion}
\label{sec:discussion}
In summary, we have demonstrated that the continuous spectrum of an ideal MHD fluid possesses nontrivial topology.  Because the continuous spectrum is associated with singular eigenfunctions, the 3D plasma breaks up into layers of 2D magnetic surfaces that are considered individually.  Regularization was required to deal with the improper eigenfunctions.  In the presence of magnetic shear, we have found a nonzero Chern number of the {\Alfven} continuum on a given magnetic surface.  The Chern number depends on the sign of the magnetic shear.  By appealing to the bulk-boundary correspondence, modes are expected to occur in a plasma where the shear changes sign.  Such modes are confirmed numerically.  This result adds to the growing understanding of the role of topology in fluids.

Our work therefore suggests that these reversed-shear {\Alfven} eigenmodes, which have been observed in tokamaks, may have topological character.  An intriguing possibility raised by this work is that RSAEs may be topologically robust in the presence of three-dimensional perturbations such as magnetic islands.  This is far from certain, however, and further investigation of RSAEs and topological phenomena is needed.  Full consideration of the details of toroidal geometry, such as the toroidicity-induced band gaps in the {\Alfven} continuum spectrum in which the RSAEs reside, are left for future work.  More broadly, this work raises the possibility that fusion devices might be carefully tailored to produce other topological states with potentially useful behavior.

These findings also provide potential research directions in astrophysics and space physics.  {\Alfven} waves can play an important role in solar physics \cite{mathioudakis:2013, belien:1996}, and whether manifestations of topological physics are important in that setting is an unresolved question.  Additionally, whistler waves guided by magnetic fields commonly occur in the ionosphere of Earth and other planets.  In attempting to understand how microscopic physics might regularize MHD, we were motivated to consider electron MHD, which ignores ion dynamics and focuses on electrons.  Whistler waves are contained within electron MHD, and in Sec.~\ref{sec:hallmhd} we showed that within electron MHD the whistler band has nontrivial topological phase.  Further investigations of the whistler band could prove fruitful.

\begin{acknowledgments}
We thank Ian Abel, Vinicius Duarte, Nikolai Gorelenkov, and Xian-Zhu Tang for useful discussions.  JBP and JBM would like to acknowledge the workshop ``Vorticity in the Universe'' held at the Aspen Center for Physics in the summer of 2017 and supported by National Science Foundation Grant PHY-1607611, which played an important role in bringing about this work.  J.B.P.'s work was performed under the auspices of the U.S.\ Department of Energy by Lawrence Livermore National Laboratory under Contract No.\ DE-AC52-07NA27344.  J.W.B.'s work was supported by the Los Alamos National Laboratory LDRD program under Project No.~20180756PRD4.  S.M.T.\ is supported by the European Research Council (ERC) under the European Union Horizon 2020 research and innovation program (Grant Agreement No.~D5S-DLV-786780).  J.B.M.~and S.M.T.\ are supported in part by a grant from the Simons Foundation (Grant No.\ 662962, GF).
\end{acknowledgments}

\appendix

\section{Ideal Nonlinear and Linear MHD}
\label{app:idealmhd}
In this section, we briefly review the most relevant facets of MHD.  The nonlinear ideal MHD equations are given by the continuity equation, the momentum equation, the energy equation, and the induction equation:
	\begin{gather}
		\pd{\r}{t} + \nabla \cdot (\r \v{v}) = 0, \label{app:MHDcontinuity} \\
		\r\left( \pd{\v{v}}{t} + \v{v} \cdot \nabla \v{v} \right) = -\nabla p + \v{J} \times \v{B} - \r\nabla \Phi, \\
		\pd{p}{t} + \v{v} \cdot \nabla p + \gamma p \nabla \cdot \v{v} = 0, \\
		\pd{\v{B}}{t} = \nabla \times (\v{v} \times \v{B}). \label{app:MHDinduction}
	\end{gather}
Here, $\Phi$ is an external gravitational potential and $\v{J} = \nabla \times \v{B}$ is the current density (units are taken such that the permeability of free space $\m_0$ is equal to 1).

These equations can be linearized around an inhomogeneous plasma equilibrium to obtain the equations determining the behavior of perturbations.  We assume the equilibrium is at rest.  The equilibrium condition is determined by force balance,
	\begin{equation}
		-\nabla p + \v{J} \times \v{B} -\r \nabla \Phi = 0.
		\label{app:force_balance}
	\end{equation}

The linearized MHD equations about a static equilibrium are
\begin{gather}
		\pd{\r_1}{t} = -\nabla \cdot (\r_0 \v{v}_1), \label{app:linearizedMHDcontinuity} \\
		\r_0 \pd{\v{v}_1}{t} = -\nabla p_1 + (\nabla \times \v{B}_1) \times \v{B}_0  \notag \\
				\qquad \qquad +  (\nabla \times \v{B}_0) \times \v{B}_1 - \r_1 \nabla \P, \\
		\pd{p_1}{t} = -\v{v}_1 \cdot \nabla p_0  -  \g p_0 \nabla \cdot \v{v}_1, \\
		\pd{\v{B}_1}{t} = \nabla \times (\v{v}_1 \times \v{B}_0). \label{app:linearizedMHDinduction}
	\end{gather}
It is possible to write all of the perturbed quantities $\r_1$, $\v{v}_1$, $\p_1$, and $\v{B}_1$ in terms of the fluid displacement vector $\bm{\xi}$, which satisfies $\partial \bm{\xi}/ \partial t = \v{v}_1$.  The linearized ideal MHD equation for $\bm{\xi}$ is given by
	\begin{equation}
		\r \pdd{\bm{\xi}}{t} = \v{F}(\bm{\xi}),
		\label{app:idealmhd_xiequation1}
	\end{equation}
where
	\begin{align}
		\v{F}(\bm{\xi}) &= -\nabla p_1  -  \v{B} \times (\nabla \times \v{B}_1)  \notag \\
		& \quad + (\nabla \times \v{B}) \times \v{B}_1  + (\nabla \Phi) \nabla \cdot (\r \bm{\xi}),
	\end{align}
the quantities $\v{B}$, $p$, and $\r$ without subscripts refer to the equilibrium variables, and the perturbed magnetic field and pressure are given in terms of $\bm{\xi}$ by
	\begin{align}
		\v{B}_1 &= \nabla \times (\bm{\xi} \times \v{B}) \\
		p_1 &= -\g p \nabla \cdot \bm{\xi}  - \bm{\xi} \cdot \nabla p.
	\end{align}

It is a celebrated result of MHD theory that the force operator $\v{F}$ is self-adjoint \cite{bernstein:1958}.  That is, given any two allowed displacements $\bm{\xi}$ and $\bm{\eta}$,
	\begin{equation}
		\int d\v{x}\, \bm{\eta}^* \cdot \v{F}( \bm{\xi} )  =  \int d\v{x}\, \bm{\xi}^* \cdot \v{F}( \bm{\eta}).
		\label{app_idealmhd_selfadjointness}
	\end{equation}

If one assumes normal modes with oscillatory dependence, $\bm{\xi} \sim e^{-i\w t}$, then Eq.~\eref{app:idealmhd_xiequation1} becomes
	\begin{equation}
		-\r \w^2 \bm{\xi} = \v{F}(\bm{\xi})
		\label{app_idealmhd_xiequation2}
	\end{equation}
The self-adjointness of $\v{F}$ guarantees that $\w^2$ is real.

\section{Linearized MHD equations in a slab}
\label{app:linearizedmhdslab}
We consider an inhomogeneous slab, where equilibrium quantities depend only on $x$.  The magnetic field is $\v{B}(x) = B_y(x) \unit{y} + B_z(x) \unit{z}$, such that magnetic surfaces are planes of constant $x$.  The acceleration due to gravity is taken to be $-\hat{g}\unit{x}$, where $\hat{g}$ is constant.  The force balance condition of Eq.~\eref{app:force_balance} becomes
	\begin{equation}
		\d{}{x} \left( p + \frac{1}{2} B^2 \right) = -\r \hat{g}.
	\end{equation}

Because the equilibrium is homogeneous in $y$ and $z$, we may Fourier transform the perturbations such that $\bm{\xi}(\v{r}) = \bm{\xi}(x) e^{i(k_y y + k_z z)}$.  Rather than using the Cartesian components $(\xi_x, \xi_y, \xi_z)$, it is more natural to use a projection onto vectors defined relative to the magnetic field.  We project onto the position-dependent, orthogonal basis $\bigl(\unit{x}, \unit{e}_\perp(x), \unit{b}(x)\bigr)$, where $\unit{b}(x) = \v{B} / B$ and $\unit{e}_\perp(x) = \unit{b} \times \unit{x}$.  We define
	\begin{subequations}
	\begin{align}
		\xi(x) &\defineas \bm{\xi} \cdot \unit{x}, \\
		\eta(x) &\defineas \bm{\xi} \cdot \unit{e}_\perp, \\
		\z(x) &\defineas \bm{\xi} \cdot \unit{b}.
	\end{align}
	\end{subequations}
In this projection, Eq.~\eref{app_idealmhd_xiequation2} becomes
	\begin{widetext}
	\begin{subequations}
	\label{app:full_spectral_matrix_equation}
	\begin{align}
		-\r \w^2 \xi &= [(\g p + B^2) \xi']' - f^2 B^2 \xi + [g(\g p + B^2) \eta]' + \r g \hat{g} \eta + (f\g p \z)' + \r f \hat{g} \z, \\
		-\r \w^2 \eta &= -g(\g p + B^2) \xi'  + \r g \hat{g} \xi - [g^2(\g p + B^2) + f^2 B^2 ]\eta - g f \g p \z, \\
		-\r \w^2 \z &= -f \g p \xi'  +  \r f \hat{g} \xi  - f g \g p \eta  - f^2 \g p \z,
	\end{align}
	\end{subequations}
	\end{widetext}
where a prime denotes a derivative with respect to $x$ and
	\begin{align}
		f &\defineas (k_y B_y + k_z B_z) / B, \\
		g &\defineas (k_y B_z - k_z B_y) / B.
	\end{align}
are the parallel and the signed perpendicular wavenumbers.  For a complete derivation of this wave equation, see Ref.~\cite{goedbloed:text}.

Equation~\eref{spectral_matrix_equation} is obtained from Eq.~\eref{app:full_spectral_matrix_equation} by making the low-plasma-beta approximation.  In other words, the plasma pressure is assumed to be much smaller than the magnetic pressure, or equivalently, the sound speed is assumed much smaller than the {\Alfven} speed.

\section{Frobenius series}
\label{app:frobenius}
In this section, we will determine the Frobenius series solution to the differential equation \eref{xi_differential_equation} for $\xi$ to find the leading-order behavior.  We consider a particular surface $x_0$ and define $s = x - x_0$.  We rewrite Eq.~\eref{xi_differential_equation} as
	\begin{equation}
		\xi'' + \frac{p(s)}{s} \xi' - \frac{q(s)}{s^2} \xi = 0,
		\label{app:xiequationA}
	\end{equation}
where 
	\begin{align}
		p(s) &\defineas \frac{s P'(s)}{P(s)}, \label{app:def_p} \\
		q(s) &\defineas \frac{s^2 Q(s)}{P(s)}, \label{app:def_q}
	\end{align}
and a prime denotes a derivative with respect to $s$.  Because $P$ is proportional to $\w^2 - \w_A^2$, the differential equation has a regular singular point at a location when $\w_A^2 = \w^2$.  We assume that $k_y$, $k_z$, and $\w^2$ are chosen such that $\w_A^2 = \w^2$ at $s=0$.  We are concerned with the behavior near $s=0$.  For a given $k_y$ and $k_z$, we assume $(\w_A^2)'_0 \neq 0$, where a 0 subscript on any quantity means that it is evaluated at $s=0$.  We can expand
	\begin{equation}
		\w^2 - \w_A^2 = -(\w_A^2)'_0 s + O(s^2).
	\end{equation}
Substituting this expression and $\w^2 = (\w_A^2)_0 = f_0^2 v_{A0}^2$ into Eq.~\eref{xi_eqn_P}, we find
	\begin{equation}
		P(s) = \frac{\r_0 (\w_A^2)'_0}{g_0^2}s + O(s^2).
	\end{equation}
Similarly, from Eq.~\eref{xi_eqn_Q} we obtain
	\begin{equation}
		Q(s) = Q_0 + O(s),
	\end{equation}
where
	\begin{align}
		Q_0 &= Q_{0a} + Q_{0b}, \\
		Q_{0a} &\defineas -\r_0' \hat{g}, \\
		Q_{0b} &\defineas \frac{\r_0 \hat{g}}{g_0^2 v_{A0}^2} (\w_A^2)'_0.
	\end{align}
Splitting $Q_0$ into two terms will be useful later.  Substituting these expressions into Eqs.~\eref{app:def_p} and \eref{app:def_q}, we obtain
	\begin{align}
		p(s) &= 1 + O(s), \\
		q(s) &= \frac{g_0^2 Q_0}{\r_0 (\w_A^2)'_0} s + O(s^2),
	\end{align}
confirming that $p(s)$ and $q(s)$ are analytic at $s=0$.

We use the method of Frobenius to obtain a solution to Eq.~\eref{app:xiequationA} in the neighborhood of $s=0$.  We assume a series
	\begin{equation}
		\xi(s) = \sum_{n=0}^\infty u_n s^{\n + n}.
	\end{equation}
and also let $p = \sum_n p_n s^n$ and $q = \sum_n q_n s^n$.  Using these series in Eq.~\eref{app:xiequationA}, and that $p_0 = 1$ and $q_0 = 0$, then balancing powers of $s$, we obtain the two lowest-order equations,
	\begin{gather}
		\n^2 u_0 = 0, \\
		\bigl[ \n(\n+1) + \n+1 \bigr] u_1 = q_1 u_0. \label{app:frobenius_u1eqn}
	\end{gather}
Hence, we have the repeated indicial exponent $\n=0$.  One solution is therefore given by
	\begin{equation}
		\xi_1(s) = u(s) = 1 + u_1 s + O(s^2),
	\end{equation}
where we have set $u_0$ to 1.  Moreover, Eq.~\eref{app:frobenius_u1eqn} determines $u_1$ as
	\begin{equation}
		u_1 = q_1 = \frac{g_0^2 Q_0}{\r_0 (\w_A^2)'_0}.
	\end{equation}
As with $Q_0$, it is convenient to split $u_1$ into the sum of two terms, and we have
	\begin{equation}
		u_1 = u_{1a} + u_{1b},
	\end{equation}
where
	\begin{align}
		u_{1a} &\defineas \frac{g_0^2 Q_{0a}}{\r_0 (\w_A^2)'_0}, \\
		u_{1b} &\defineas \frac{\hat{g}}{v_{A0}^2}.
	\end{align}
Because of the repeated indicial exponent, a second linearly independent solution has the form
	\begin{equation}
		\xi_2(s) = u(s) \ln s + v(s),
	\end{equation}
where $v(s)$ is analytic at $s=0$.  The logarithmic singularity in $\xi_2$ dominates the behavior near $s=0$, and we will not need any terms in the series for $u$ beyond $u_1$.  The topology of the magnetic surface at $s=0$ is determined by $u_{1a}$, and all higher-order terms are subdominant, as are all the terms in $v(s)$.

\section{Evaluating the Chern integral}
\label{app:integral}
In this section, we evaluate the Chern integral of Eq.~\eref{Chern_integral}, restated here:
	\begin{equation}
	C = \frac{1}{2\pi} \int dk_y\, dk_z\, \mathcal{F}_x.
	\label{app:Chern_integral}
	\end{equation}
This entails integrating the Berry curvature $\mathcal{F}_x$ [given in Eq.~\eref{Berrycurvature}] over the $(k_y, k_z)$ plane to obtain the Chern number.  We work in polar coordinates with $k_y = k \cos \vp$ and $k_z = k \sin \vp$.  We observe that the only $k$ dependence in $\mathcal{F}_x$ occurs within $w$ [given in Eq.~\eref{definition_of_w}], which has the form $w = w_1 + w_2 k^2$, with
	\begin{align}
		w_1 &=  u_{1a}^2 \ln^2 \e  +  \left( \frac{1}{\e} - \frac{\pi u_{1a}}{2} \right)^2 + \frac{\cos^2\vp}{\sin^2 \vp} \frac{\hat{g}^2}{v_{A0}^4} \ln^2 \e, \\
		w_2 &= \cos^2 \vp \ln^2 \e.
	\end{align}
The integral over $k$ is straightforward,
	\begin{equation}
		\int_0^\infty dk \frac{k}{w^2} = \frac{1}{2 w_1 w_2}.
	\end{equation}
What remains is the angle integral.  To make the manipulations more convenient, we define the following quantities:
	\begin{align}
		n_1 &\defineas \frac{Q_{0a}}{B_0^2}, \label{app:defn1} \\
		n_2^2 &\defineas \frac{\hat{g}^2}{ v_{A0}^4}, \label{app:defn2} \\
		m_1 &\defineas \frac{2}{L_s}, \\
		m_2 &\defineas \frac{1}{L_{\r B}}, \label{app:defm2} \\
		\g(\vp) &\defineas m_1 \cos \vp \sin \vp - m_2 \sin^2 \vp.
	\end{align}
After some algebra, we find that \eref{app:Chern_integral} becomes
	\begin{equation}
		C  = -\frac{\e \ln \e}{2\pi} n_1 I,
		\label{app:Chern_integral_with_I}
	\end{equation}
where
	\begin{widetext}
	\begin{equation}
		I = \int_0^{2\pi} d\vp \frac{ m_1 \cos^2 \vp - 2m_2 \cos \vp \sin \vp}{\bigl[ n_1^2 \cos^4 \vp + n_2^2 \cos^2 \vp (m_1 \cos \vp - m_2 \sin \vp)^2 \bigr] \e^2 \ln^2 \e + \bigl(\g - \frac{1}{2} \e \pi n_1 \cos^2 \vp\bigr)^2}.
		\label{app:Chern_phi_integral}
	\end{equation}
	\end{widetext}
We now evaluate $I$  as a function of $n_1, n_2, m_1$, and $m_2$ in the limit $\e \to 0$.  In this limit, the $\e \pi n_1 \cos^2 \vp$ term in the denominator can be neglected.

For small $\e$, the dominant contribution to $I$ arises near the angles where $\g(\vp)$ vanishes, which occurs when $\sin\vp = 0$ and $\tan\vp= m_1 / m_2$.  For example, see the plot of the Berry curvature in Fig.~\ref{fig:berry_curvature}(b).  Near the resonance peaks where $\g(\vp)$ vanishes, the integrand can be approximated as a Lorentzian.

The angles that provide the dominant contribution are the same as those in which $(\omega_A^2)'_0 = 0$ and the Frobenius solution of Eq.~\eqref{xi2_singular} breaks down; see Fig.~\ref{fig:berry_curvature}(a).  However, these lines on which the Frobenius solution is invalid are points of measure zero in the area integral and do not pose an impediment to evaluating the integral.  As pointed out previously, the Berry curvature is finite and continuous.

Let us first consider the resonance peaks where $\sin\vp =0$.  We calculate the contribution due to the region around $\vp=0$; the contribution from near $\vp=\pi$ is identical.  Near $\vp=0$, we approximate $\sin \vp \approx \vp$ and $\cos \vp \approx 1$.  Keeping the dominant terms, the contribution from $\vp$ near zero is given by
	\begin{equation}
		I_1 = m_1 \int_{-\D}^{\D}  d\vp \frac{1}{(n_1^2 + n_2^2 m_1^2) \e^2 \ln^2 \e  +  m_1^2 \vp^2}
	\end{equation}
for some width $\D$ around $\vp = 0$.  In this form, we can extend the limits on the integral to infinity and use the Lorentzian integral
	\begin{equation}
		\int_{-\infty}^\infty dx \frac{1}{a_1^2 + a_2^2 x^2} = \frac{\pi}{|a_1 a_2|}.
	\end{equation}
Hence, we find
	\begin{equation}
		I_1 = \frac{\pi}{\e |\ln \e|}  \frac{\sgn(m_1)}{\sqrt{n_1^2 + n_2^2 m_1^2}}.
	\end{equation}

Next, we consider the contribution to $I$ from the resonance peaks where $\tan \vp = m_1/m_2$.  We consider the contribution $I_2$ from the region around $\vp_0 = \tan^{-1}(m_1/m_2)$; the contribution from $\vp_0 + \pi$ is identical.  It is convenient to rewrite the integrand by dividing numerator and denominator by $\cos^4 \vp$, and we have
	\begin{widetext}
	\begin{equation}
		I_2 = \int_{\vp_0 - \D}^{\vp_0 + \D} d\vp \frac{\sec^2\vp (m_1 - 2m_2 \tan \vp)}{[n_1^2 + n_2^2(m_1 - m_2 \tan \vp)^2 ] \e^2 \ln^2 \e  +  \tan^2 \vp (m_1 - m_2 \tan \vp)^2}.
	\end{equation}
	\end{widetext}

Since $\tan \vp_0 = m_1/m_2$, we note
	\begin{align}
		\sin \vp_0 &= \frac{m_1}{\sqrt{m_1^2 + m_2^2}}, \\
		\cos \vp_0 &= \frac{m_2}{\sqrt{m_1^2 + m_2^2}}.
	\end{align}
For nonzero $m_1$, this resonance is well separated from the other resonance at $\vp=0$  in the limit $\e \to 0$.  Near $\vp \approx \vp_0$, we may use $\sec^2 \vp \approx \sec^2 \vp_0 = (m_1^2 + m_2^2) / m_2^2$ and $m_1 - 2 m_2 \tan \vp \approx m_1 - 2 m_2 \tan \vp_0 = -m_1$.  We approximate
	\begin{equation}
		m_1 - m_2 \tan \vp \approx - \frac{m_1^2 + m_2^2}{m_2} (\vp - \vp_0).
	\end{equation}
Substituting these expressions, changing integration variables to $x = \vp - \vp_0$, and then extending the limits of the integral to $\pm \infty$, we obtain
	\begin{equation}
		I_2 = - \int_{-\infty}^\infty dx\, \frac{m_1 (m_1^2 + m_2^2) / m_2^2}{n_1^2 \e^2 \ln^2 \e +  m_1^2 (m_1^2 + m_2^2)^2 x^2 / m_2^4},
	\end{equation}
where an $x^2 \e^2$ term in the denominator has been neglected as subdominant.  This Lorentzian integral evaluates to
	\begin{equation}
		I_2 = -\frac{\pi}{\e |\ln \e|} \frac{\sgn(m_1)}{|n_1|} .
	\end{equation}
Hence,
	\begin{equation}
		I = 2(I_1 + I_2) = -\frac{2\pi}{\e |\ln \e|} \frac{\sgn(m_1)}{|n_1|} \left( 1 - \frac{1}{\sqrt{1 + \frac{n_2^2 m_1^2}{n_1^2}}} \right).
		\label{app:integral_I_general}
	\end{equation}

In this evaluation of $I$, we have left $n_1$, $n_2$, $m_1$, and $m_2$ as general parameters.  However, the definitions in Eqs.~\eref{app:defn1}--\eref{app:defm2} imply $n_1^2 / n_2^2 = m_2^2$.  Thus,			\begin{equation}
		I = -\frac{2\pi}{\e |\ln \e|} \frac{\sgn(m_1)}{|n_1|} \left( 1 - \frac{1}{\sqrt{1 + \frac{m_1^2}{m_2^2}}} \right).
	\end{equation}
Substituting this expression into \eref{app:Chern_integral_with_I}, we find that in the limit $\e \to 0$, the Chern number associated with a given magnetic surface is
	\begin{equation}
		C = -\sgn(\r_0' \hat{g} L_s) \left(1 - \frac{1}{\sqrt{1 + \frac{4 L_\r^2}{L_s^2}}} \right).
		\label{app:Chern_result}
	\end{equation}
The first term in parentheses in Eq.~\eref{app:Chern_result} arises from the $\tan\vp = m_1/m_2$ resonances, while the second term arises from the $\sin\vp = 0$ resonances.  Note that we have obtained a finite result for the Chern number in the $\e \to 0$ limit, and it was not at all obvious at the outset this would occur.  We have confirmed our analytic results by comparing Eq.~\eref{app:integral_I_general} with numerical integrations of Eq.~\eref{app:Chern_phi_integral}.  

We implicitly assumed $m_1 \neq 0$, which justified treating the resonances as separated.  If the magnetic shear vanishes, implying $m_1=0$, then instead the result is $C=0$.  This can be seen easily from Eq.~\eref{app:Chern_phi_integral}.  If $m_1=0$, then the numerator of the integrand is odd in $\vp$ and the denominator is even in $\vp$, so the integral over $\vp$ vanishes.

\section{Trivial topology in reduced MHD}
\label{app:reducedmhd}
Reduced MHD offers a simplified description of plasma behavior based on an asymptotic reduction that is often used in theoretical analyses \cite{schnack:2009lectures}.  It is natural to ask whether nontrivial topology can be extracted from reduced MHD, or whether full MHD is required.  We conclude that full MHD is necessary.  

One assumption of reduced MHD is low beta, a limit we have already taken.  Another assumption is that fluctuations are anisotropic, such that typical fluctuations have longer wavelengths along the magnetic field line than across it.  Equivalently, the parallel wavenumber $f$ is assumed much smaller than the perpendicular wave number $g$.  Finally, one assumes the equilibrium magnetic field is primarily unidirectional, which we take to be the $z$ direction, such that $B_z \gg B_y$.  The reduced MHD ordering contains sufficient physics to describe RSAEs.

A consequence of these assumptions is that in Eq.~\eref{spectral_matrix_equation}, $\z$ decouples from the equations for $\xi$ and $\eta$.  Equation~\eref{spectral_matrix_equation} is replaced by
	\begin{subequations}
	\begin{align}
		-\r \w^2 \xi &= (\r v_A^2 \xi')' - \r f^2 v_A^2 \xi + (\r g v_A^2 \eta)' + \r g \hat{g} \eta, \\
		-\r \w^2 \eta &= -\r g v_A^2 \xi' + \r g \hat{g} \xi - \r (f^2 + g^2) v_A^2 \eta.
	\end{align}
	\end{subequations}
which determine the fluid displacement in the directions perpendicular to the equilibrium magnetic field.  The fluid displacement $\z$ parallel to the magnetic field is typically neglected as small.  In effect, reduced MHD offers an economical description of dynamics in the plane perpendicular to the magnetic field.

To see that reduced MHD implies a trivial topology, we demonstrate that if one treats the eigenfunction as $[\xi, \eta]$, and ignores the existence of $\z$, then one finds the Berry curvature integrates to exactly zero.  This is straightforward to see.  If $\z$ is neglected, then the third term on the right-hand side of Eq.~\eref{definition_of_w}, proportional to $\hat{g}^2$, disappears.  But that term is the one that provides the dependence on $n_2$ in Eq.~\eref{app:Chern_phi_integral}.  Hence, if $\z$ is neglected, the result of the Berry curvature integral is exactly that of Eq.~\eref{app:integral_I_general} with $n_2$ equal to zero.  The result is zero.

%

\end{document}